\documentclass[iop]{emulateapj}
\usepackage{graphicx}

\begin{document}

\submitted{Accepted for publication in The Astronomical Journal, August 6, 2012}

\title{NEW ATLAS9 AND MARCS MODEL ATMOSPHERE GRIDS FOR THE 
APACHE POINT OBSERVATORY GALACTIC EVOLUTION EXPERIMENT (APOGEE)}

\author{
Sz.~M{\'e}sz{\'a}ros\altaffilmark{1,2}, 
C.~Allende~Prieto\altaffilmark{1,2}, 
B.~Edvardsson\altaffilmark{3}, 
F.~Castelli\altaffilmark{4}, 
A.~E.~Garc{\'{\i}}a~P{\'e}rez\altaffilmark{5}, 
B.~Gustafsson\altaffilmark{3}, 
S.~R.~Majewski\altaffilmark{5}, 
B.~Plez\altaffilmark{6}, 
R.~Schiavon\altaffilmark{7}, 
M.~Shetrone\altaffilmark{8}, 
A.~de~Vicente\altaffilmark{1,2}
}

\altaffiltext{1}{Instituto de Astrof{\'{\i}}sica de Canarias (IAC), E-38200 La Laguna, Tenerife, Spain}
\altaffiltext{2}{Departamento de Astrof{\'{\i}}sica, Universidad de La Laguna (ULL), E-38206 la Laguna, Tenerife, Spain}
\altaffiltext{3}{Department of Physics and Astronomy, Division of Astronomy and Space Physics, Box 515, 
751 20 Uppsala, Sweden}
\altaffiltext{4}{Istituto Nazionale di Astrofisica, Osservatorio Astronomico di Trieste, via Tiepolo 11, 
34143 Trieste, Italy}
\altaffiltext{5}{Department of Astronomy, University of Virginia, P.O. Box 400325, Charlottesville, VA 22904-4325, USA}
\altaffiltext{6}{Laboratoire Univers et Particules de Montpellier, Universit\'e Montpellier 2, CNRS, 34095 Montpellier, 
France}
\altaffiltext{7}{Gemini Observatory, 670 North A`ohoku Place, Hilo, HI 96720, USA}
\altaffiltext{8}{McDonald Observatory, University of Texas, Austin TX 78712, USA}

\begin{abstract}

We present a new grid of model photospheres for the SDSS-III/APOGEE
survey of stellar populations of the Galaxy, calculated using the
ATLAS9 and MARCS codes. New opacity distribution functions were generated to calculate ATLAS9 model
photospheres. MARCS models were calculated based on opacity sampling techniques. 
The metallicity ([M/H]) spans from $-$5 to 1.5 for ATLAS and $-$2.5 to 0.5 for MARCS models. 
There are three main differences with respect to  
previous ATLAS9 model grids: a new corrected H$_{\rm 2}$O linelist, 
a wide range of carbon ([C/M]) and $\alpha$ element [$\alpha$/M] 
variations, and solar reference abundances from \citet{asplund01}. The added range of varying carbon and $\alpha$ element
abundances also extends the previously calculated MARCS model grids. 
Altogether 1980 chemical compositions were used for the ATLAS9 grid, and 175 for the MARCS grid. 
Over 808 thousand ATLAS9 models were computed spanning temperatures from
3500\,K to 30000\,K and log~$g$ from 0 to 5, where larger temperatures only have high gravities. The MARCS models span
from 3500\,K to 5500\,K, and log~$g$ from 0 to 5. All model atmospheres are publically available online.

\end{abstract}

\section{Introduction}

The Apache Point Observatory Galactic Evolution Experiment (APOGEE; Allende Prieto et al. 2008) 
is a large-scale, near-infrared, high-resolution spectroscopic survey of Galactic stars, 
and it is one of the four experiments in the Sloan Digital Sky Survey-III (SDSS-III; Eisenstein et al. 2011; 
Gunn et al. 2006, Aihara et al. 2011). APOGEE will obtain high S/N, R$\sim$22,500 spectra for 100,000
stars in the Milky Way Galaxy, for which accurate chemical abundances,
radial velocities, and physical parameters will be determined.
APOGEE data will shed new light on the formation of the Milky Way,
as well as its chemical and dynamical evolution. To achieve its
science goals, APOGEE needs to determine abundances for about 15 elements to an accuracy of 0.1 dex. To attain this
precision, a large model photosphere database with up-to-date solar
abundances is required. We chose to build the majority of APOGEE's
model photosphere database on ATLAS9 and MARCS calculations. 

ATLAS \citep{kurucz05} is widely used as a universal LTE 1$-$D plane-parallel atmosphere modeling code, which is 
freely available from Robert Kurucz's website\footnote{http://kurucz.harvard.edu}. ATLAS9 \citep{kurucz01} handles the 
line opacity with the opacity distribution functions (ODF), which greatly simplifies and reduces the computation time 
\citep{strom01, kurucz04, castelli02}. ATLAS uses the mixing-length scheme for convective energy transport. 
It consists in pretabulating  the line opacity as function of temperature  and gas pressure in a given number of 
wavelength intervals which cover the whole wavelength range from far ultraviolet to far infrared. For computational 
reasons, in each interval the line opacities are rearranged according to strength rather than wavelength.
For each selected metallicity and microturbulent velocity, an opacity distribution function table has to be computed.
While the computation of the ODFs is very time consuming, extensive grids of model atmospheres and spectrophotometric 
energy distributions can be computed in a short time once the required ODF tables are available.

ATLAS12 \citep{kurucz04, castelli03} 
uses the opacity sampling method (OS) to calculate the opacity at 30,000 points. 
The high resolution synthetic spectrum at a selected resolution can then be obtained by running SYNTHE \citep{kurucz03}. 
More recently, \citet{lester01} have developed SATLAS\_\_ODF and SATLAS\_\_OS, the spherical version of both ATLAS9 and 
ATLAS12, respectively. No extensive grids of models have been published up to now, neither with ATLAS12, nor with any of 
the two versions of SATLAS. 

Instead, extensive grids of ATLAS9 ODF model atmospheres for several metallicities were calculated by 
\citet{castelli01}. These grids are based on solar (or scaled solar) abundances from \citet{grevesse01}. 
Recently, \citet{kirby01} provided a new ATLAS9 grid, but he used abundances from \citet{anders01}. The calculations 
presented 
in this paper are based on the more recent solar composition from \citet{asplund01}. This updated abundance table 
required new ODFs and Rosseland mean 
opacity calculations as well. Abundances from \citet{asplund01} were chosen instead of those from newer studies 
\citep{asplund02} to match the composition of the MARCS models described below, and those available from the MARCS website. 

The MARCS model atmospheres \citep{gustafsson01, plez01, gustafsson02} were developed 
and have been evolving in close connection with applications primarily to spectroscopic analyses of a wide range 
of late type stars with different properties. The models are one-dimensional plane-parallel or spherical, and computed 
in LTE assuming the mixing-length scheme for convective energy transport, as formulated by \citet{henyey01}. For 
luminous stars (giants), where the geometric depth of the photosphere is a non-negligible fraction of the stellar 
radius, the effects of the radial dilution of the energy transport and the depth-varying gravitational field 
is taken into account. Initially, spectral line opacities were economically 
treated by the ODF approximation, but later the more flexible and realistic opacity 
sampling scheme has been adopted. In the OS scheme, line opacities are directly 
tabulated for a large number of wavelength points (10$^{5}$) as a function of temperature and
pressure. 

The shift in the MARCS code from using ODFs to the OS scheme avoided the sometimes 
unrealistic assumption that the line opacities of certain relative strengths within each ODF wavelength interval overlap 
in wavelength irrespective of depth in the stellar atmosphere. This assumption was found to lead to systematically 
erroneous models, in particular when plolyatomic molecules add important opacities to surface layers \citep{ekberg01}. 
The current version of the MARCS code used for the present project and for the more extensive MARCS model atmosphere data 
base\footnote{http://marcs.astro.uu.se/} was presented and described in detail by \citet{gustafsson02}. 
The model atmospheres presented in this paper adds large variaty in [C/M] and [$\alpha$/M] abundances to the 
already existing grids by covering these abundances systematically from -1 to +1 for each metallicity. 

Our main purpose is to update the previous ATLAS9 grid and publish new MARCS models to provide a large 
composition range to use in the APOGEE survey and future precise abundance analysis projects. These new ATLAS models were 
calculated with a corrected H$_{2}$O linelist. The abundances used for the
MARCS models presented in this paper are from \citet{grevesse02}, which are nearly identical to \citet{asplund01}; 
the only significant difference is an abundance of scandium 0.12 dex higher than in \citet{asplund01}. The range of
stellar parameters (T$_{\rm eff}$, log~$g$ and [M/H]) spanned by 
the models covers most stellar types found in the Milky Way.

This paper is organized as follows. In Section 2 we describe the parameter range of
our ODFs and model atmospheres and give details of the calculation method of ATLAS9 we implemented. Section 3 contains
the parameter range and calculation procedure for MARCS models. 
In Section 4 we compare MARCS and ATLAS9 models with \citet{castelli01}, and illustrate how different 
C and $\alpha$ contents affect the atmosphere. Section 5 contains the conclusions. 
The grid of ODFs and model atmospheres will be periodically updated in the future and available 
online\footnote{http://www.iac.es/proyecto/ATLAS-APOGEE/}.

\begin{figure}
\includegraphics[width=2.5in,angle=270]{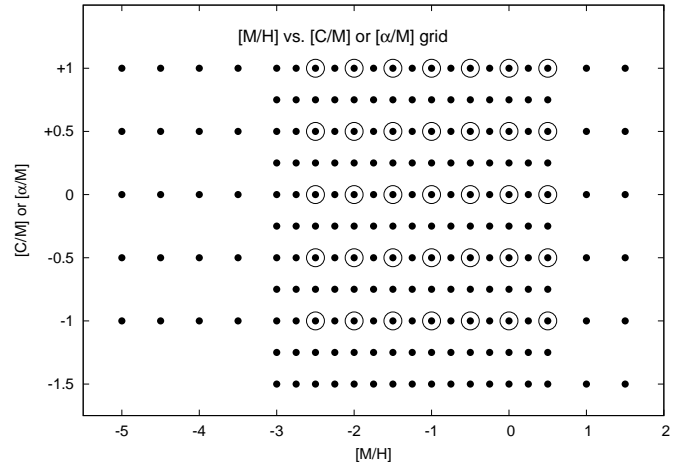}
\caption{The [C/M] or [$\alpha$/M] content as a function of [M/H] of ATLAS9 models (filled circles, Table 1) and MARCS
models (open circles, Table 4). Both [C/M] and 
[$\alpha$/M] changes independently from each other, and the small steps in 
metallicities give altogether 1980 different compositions for the ATLAS9 models and 175 compositions for the MARCS models. 
The number of acceptable models may vary for each composition, for details see the ATLAS-APOGEE 
website (http://www.iac.es/proyecto/ATLAS-APOGEE/). For missing 
metal-rich compositions of ATLAS9 models see Table 2.}
\end{figure}

\section{ATLAS9 Model Atmospheres}

\subsection{Parameters}

The metallicity ([M/H]) of the grid varies from $-$5 to 1.5 to cover the full range of chemical 
compositions and scaled to solar abundances\footnote{[M/H] means any element with Z $>$ 2 and  
[M/H] = $\log_{10}(N_{\rm M} / N_{\rm H})_{\star} - 
\log_{10}(N_{\rm M} / N_{\rm H})_{\odot}$, where N$_{\rm Fe}$ and N$_{\rm H}$ are the number of the desired element 
and hydrogen nuclei per unit volume, respectively}. For each of these solar scaled compositions we also vary the [C/M] 
and [$\alpha$/M] abundances from $-$1.5 to 1 (Figure 1).

ODFs and Rosseland opacity files were calculated with microturbulent velocities 
v$_{\rm t}$ = 0, 1, 2, 4, 8 km~s$^{-1}$, while the model atmospheres were produced only with 
v$_{\rm t}$ = 2 km~s$^{-1}$. 
The metallicity grids were the same for all effective temperatures, and the 
range can be seen in Table 1. Some metal-rich compositions with high C but low $\alpha$ content were not calculated due
to excessive computation time; these are listed in Table 2. 
The $\alpha$ elements considered when varying [$\alpha$/M] were the
following: O, Ne, Mg, Si, S, Ca, and Ti. 
The temperature and gravity parameter grid for each composition and spectral type is given in Table 3. 
The T$_{\rm eff}$ $-$ log~$g$ distribution is plotted in Figure 2. Extreme metal-poor and metal-rich compositions 
were also included.

\begin{figure}
\includegraphics[width=2.5in,angle=270]{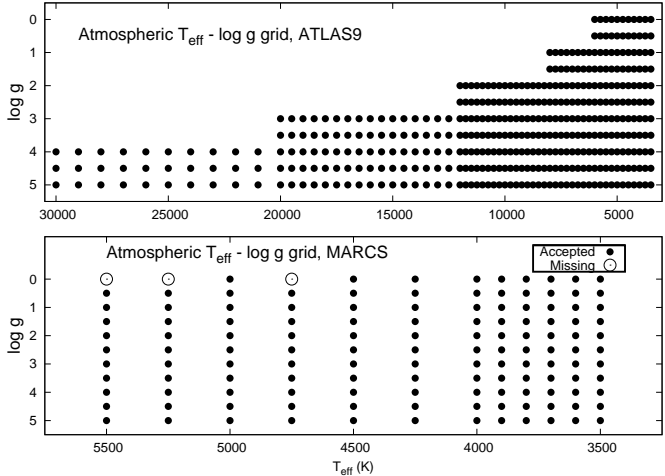}
\caption{The gravity (log~$g$) as a function of effective temperature (T$_{\rm eff}$) of ATLAS9 and MARCS  models 
calculated for each composition. Acceptable models are denoted by filled circles, while models not acceptable (and
missing) are denoted by open circles for the solar composition. Models with C/O~$>$~1.7 and T$_{\rm eff} < 4000$\,K 
are not published (see Section 4.2).
}
\end{figure}

All the ATLAS codes use atomic and molecular linelists made available by Kurucz on a series of CDROM \citep{kurucz02}. 
They can now be found at the Kurucz website\footnote{http://kurucz.harvard.edu/LINELIST/}.
The molecular line lists for TiO 
and H$_{2}$O were provided by \citet{schwenke01} and \citet{partridge01}, respectively, and reformatted by Kurucz in 
ATLAS format. These are also available for download at the Kurucz website.
For these models, we used the same linelists as \citet{castelli01}, except for H$_{2}$O, for which a new Kurucz release of 
the \citet{partridge01} data was adopted\footnote{http://kurucz.harvard.edu/MOLECULES/H2O/h2ofastfix.bin}.

The solar reference 
abundance table was adopted from \citet{asplund01}. Convection was turned on with the mixing length parameter set to 
$l/H_{p} = 1.25$, but the convective
overshooting was turned off. All the models have the same 72 layers from log $\tau_{Ross}$ = $-$6.875
to 2, where the step is log $\tau_{Ross}$ = 0.125. These parameters remained the same as \citet{castelli01} for easy
comparison.

All computations were performed on 
the Diodo cluster at the Instituto Astrofisico de Canarias. Diodo consists of 1 master node and 19 compute nodes, for a 
total of 80 cores and 256 GB of RAM, communicating through two independent Gigabit Ethernet networks. 16 of the 
compute nodes host 2 Intel Xeon 3.20 GHz EM64T processors each, with 4 GB of RAM (2GB per core); the remaining 3
compute nodes host each 16 Intel Xeon (E7340) 2.40 GHz EM64T processors, with 64 GB of RAM (4GB per core). 
On this cluster, about 3 months of computer time was required for ODF and model atmosphere calculations, using 
all 80 processors.

\subsection{Calculation Method}

Two separate scripts were developed, one for the ODF and Rosseland opacity calculations, and one for the 
ATLAS9 calculations. The ODF and Rosseland opacity 
calculations followed exactly the procedure described by \citet{castelli01}, and \citet{castelli02} using the DFSYNTHE 
code for the ODF,
KAPPA9 code for the Rosseland opacity, and ATLAS9 for the model atmopshere calculations \citep{sbordone02, sbordone01}. 
These codes were compiled in Linux with the Intel Fortran compiler version 11.1. 

\begin{deluxetable}{lrrr}
\tabletypesize{\scriptsize}
\tablecaption{Abundance Parameters of ATLAS9 models}
\tablewidth{0pt}
\tablehead{
 & \colhead{Min} & \colhead{Max}    & \colhead{Step}     }
\startdata
$[$M/H$]$ & $-$5 & $-$3.5 & 0.5 \\
$[$C/M$]$ & $-$1 & 1 & 0.5 \\
$[\alpha$/M$]$ & $-$1 & 1 & 0.5 \\
\hline
$[$M/H$]$ & $-$3 & 0.5 & 0.25 \\
$[$C/M$]$ & $-$1.5 & 1 & 0.25 \\
$[\alpha$/M$]$ & $-$1.5 & 1 & 0.25 \\
\hline
$[$M/H$]$ & 1 & 1.5 & 0.5 \\
$[$C/M$]$ & $-$1.5 & 1 & 0.5 \\
$[\alpha$/M$]$ & $-$1.5 & 1 & 0.5 \\
\enddata
\end{deluxetable}

\begin{deluxetable}{rrr}
\tabletypesize{\scriptsize}
\tablecaption{Missing compositions of ATLAS9 models}
\tablewidth{0pt}
\tablehead{
\colhead{[M/H]} & \colhead{[C/M]}    & \colhead{[$\alpha$/M]}     }
\startdata
1 & 1 & $-$1.5 \\
1 & 1 & $-$1 \\
1.5 & 0.5 & $-$1.5 \\
1.5 & 1 & $-$1.5 \\
1.5 & 1 & $-$1 \\
1.5 & 1 & $-$0.5 \\
1.5 & 1 & 0 \\
\enddata
\end{deluxetable}

\begin{deluxetable}{lrrrrrr}
\tabletypesize{\scriptsize}
\tablecaption{Model Atmosphere Parameters of ATLAS9 models}
\tablewidth{0pt}
\tablehead{
\colhead{Spectral Type} &
\colhead{$T_{\rm eff}$}    & \colhead{$T_{\rm eff}$} & \colhead{$T_{\rm eff}$}    & 
\colhead{log~$g$} & \colhead{log~$g$} & \colhead{log~$g$} \\ 
\colhead{} & 
\colhead{min}    & \colhead{max} & \colhead{step}    & 
\colhead{min} & \colhead{max} & \colhead{step}}
\startdata
M, N, R, K, G & 3500 & 6000 & 250 & 0 & 5 & 0.5 \\
F & 6250 & 8000 & 250 & 1 & 5 & 0.5 \\
A & 8250 & 12000 & 250 & 2 & 5 & 0.5 \\
B & 12500 & 20000 & 500 & 3 & 5 & 0.5 \\
B, O & 21000 & 30000 & 1000 & 4 & 5 & 0.5 \\
\enddata
\end{deluxetable}

Our algorithm sets up the initial starting models from the grid
provided by \citet{castelli01}\footnote{http://wwwuser.oat.ts.astro.it/castelli/}. The algorithm chooses 
the model that has the closest composition, effective
temperature, and log~$g$ to the desired output, and an initial ATLAS9 
model is calculated. The result must be checked to see whether the output model 
satisfies the convergence parameters provided by the user for each layer in the model atmosphere. 
These parameters were set to 1$\%$ for the flux or
10$\%$ for the flux derivative errors after 30 iterations in each run, as recommended in the ATLAS 
cookbook\footnote{http://atmos.obspm.fr/index.php/documentation}. A model is considered converged if the convergence 
parameters satisfy these criteria in all depths. 

We then determined that an atmospheric model is acceptable if one of the following criteria is satisfied: 1. the model 
has converged through the whole atmosphere, 2. no more than 1 non-converged layer exists between 
log~$\tau_{\rm Ross} = -4$ and log~$\tau_{\rm Ross} = 1$. The model is allowed to have other non-converged layers 
for log~$\tau_{\rm Ross} < -4$. A model was considered unacceptable in all other cases. 
We used only log~$\tau_{\rm Ross} >= -4$ to log~$\tau_{\rm Ross} = 1$ to check the convergence, because most of the 
lines in the optical and H band form in this region. In case the output was not acceptable, we restarted the
calculation using more iterations. In case of a run with unacceptable output, 
we selected a starting model that had a different log~$g$ from the initial starting model and used it 
to restart the calculation. Then the previously described convergence test was performed
and more restarts were done if it was necessary. If the output remained unacceptable, the effective 
temperature of the output model was changed by 10, 50, 
and 100\,K. If the output of any of these runs was acceptable, then it was used as an 
input model to calculate the atmosphere with the original effective temperature. 

To test our model atmospheres, we replicated the Rosseland opacity calculations by \citet{castelli01}. 
For this we used Castelli's scripts without any modification to calculate the ODF, and Rosseland opacities for the 
\citet{grevesse01} abundances. These calculations concluded in perfect agreement within numerical precision. 
We then attempted to reproduce model atmospheres with the same parameters found on Castelli's website with our 
scripts using the ODFs and Rosseland opacity files generated with \citet{grevesse01} abundances. This test also 
concluded with near perfect agreement with only 0.1$-$0.2\,K maximum differences coming from the different version 
of compilers used. 

\section{MARCS Model Atmospheres}

\subsection{Parameters}

Models were computed for seven overall metallicities, [M/H] from $-$2.5 to 0.5, with a step size of 
0.5 dex. For each of these seven overall [M/H] mixtures, 25 combinations of modified carbon and $\alpha$ element 
abundances were adopted: The modifications to the logarithmic C and $\alpha$ abundances are $-$1, $-$0.5, 0, 0.5, and 
1\, dex. This format resulted in a total of 175 subgrids with unique chemical compositions (Table 4). 
The $\alpha$ elements in MARCS are O, Ne, Mg, Si, S, Ar, Ca, and Ti. This composition scheme is exactly the same as 
the previous models found on the MARCS website. The systematic $\alpha$ abundance changes were chosen to overlap the 
scheme (Section 2.1) used in the ATLAS9 model calculations. 

\begin{deluxetable}{lrrr}
\tabletypesize{\scriptsize}
\tablecaption{Abundance Parameters of MARCS models}
\tablewidth{0pt}
\tablehead{
 & \colhead{Min} & \colhead{Max}    & \colhead{Step}     }
\startdata
$[$M/H$]$ & $-$2.5 & 0.5 & 0.5 \\
$[$C/M$]$ & $-$1 & 1 & 0.5 \\
$[\alpha$/M$]$& $-$1 & 1 & 0.5 \\
\enddata
\end{deluxetable}

\begin{deluxetable}{lrrrrrr}
\tabletypesize{\scriptsize}
\tablecaption{Model Atmosphere Parameters of MARCS models}
\tablewidth{0pt}
\tablehead{
\colhead{Spectral Type} &
\colhead{$T_{\rm eff}$}    & \colhead{$T_{\rm eff}$} & \colhead{$T_{\rm eff}$}    & 
\colhead{log~$g$} & \colhead{log~$g$} & \colhead{log~$g$} \\
\colhead{} &
\colhead{min}    & \colhead{max} & \colhead{step}    & 
\colhead{min} & \colhead{max} & \colhead{step}}
\startdata
M, N, R, K \tablenotemark{a} & 3500 & 4000 & 100 & 0 & 3 & 0.5 \\
K, G \tablenotemark{a} & 4250 & 5500 & 250 & 0 & 3 & 0.5 \\
M, N, R, K \tablenotemark{b} & 3500 & 4000 & 100 & 3.5 & 5 & 0.5 \\
K, G \tablenotemark{b} & 4250 & 5500 & 250 & 3.5 & 5 & 0.5 \\
\enddata
\tablenotetext{a}{Spherical atmospheres}
\tablenotetext{b}{Plane-parallel atmospheres}
\end{deluxetable}

For each of these abundance subgrids, models with 12 values of effective temperature from 3500 to 5500\,K and 11 values 
of logarithmic surface gravities from 0 to 5 were computed (see Figure 2 and Table 5). Models with logarithmic surface 
gravities lower than 3.5 (giants) were computed in spherical geometry and with a microturbulence parameter of 
2~km~s$^{-1}$, while the remaining (dwarf) models adopted v$_{\rm t}$ = 1~km~s$^{-1}$ and plane-parallel geometry. 
In the end, 86\% of the 23,140 models converged satisfactorily. Convergence was particularly poor for cool dwarfs that 
are simultaneously $\alpha$ rich and carbon poor. 

Details about the atomic and molecular line lists used by the MARCS code are given by \citet{gustafsson02}.
In a number of instances, they are different from those used by the ATLAS code, as for H$_{2}$O \citep{barber01}, 
and TiO \citep{plez03}.

\subsection{Calculation Method}

Similarly to the ATLAS9 ODF calculations, a new metallicity MARCS subgrid is started with the summation of an 
OS table of atomic line opacities for the relevant abundance mixture and microturbulence parameter. Since line 
opacity data is needed for many atoms and first ions, it saves time to add the opacities of the individual species 
into one file before a set of models is computed. This file contains a table with over 10$^5$ wavelength points with line 
opacities relevant to the equation of state for 306 combinations of temperature and damping pressure $P_6$. 
The damping pressure is used as a proxy for the pressure that broadens metal lines by collisions with neutral 
atoms in cool stars. $P_6$ is the pressure of HI with the addition of the polarizability corrected pressures 
of neutral helium and H$_2$ (see Eq. 33 of Gustafsson et al. 2008). For each molecule, in contrast, the 
line opacities are given in one table for the same wavelength set, but as a function of temperature and 
microturbulence parameter.

The method used for the MARCS model calculations is described in detail in \citet{gustafsson02}. They start with the 
generation 
of a simplified starting model assuming a grey opacity. Physical parameters and their derivatives are computed 
and the model structure is iterated in a multidimensional Newton-Raphson scheme until the flux through each depth layer 
corresponds to the prescribed effective temperature. The models usually converged after 4 to 8 iterations; 
convergence also requires that the maximum temperature correction in any depth point is below 
1.5\,K during two consecutive iterations. Occasionally convergence takes longer, and some models do not converge 
at all. A converged model with similar model parameters is then identified as a new starting model. This approach is 
often successful, but some models do not converge, which leaves vacancies in the model grid.

\section{Discussion}

Changes in the chemical composition can have a number of effects on the calculated atmospheres. These effects are 
mainly related to either changes in opacity or changes in the equation of state. The main effect of an increased line
opacity in late-type stars is either cooling or warming (depending on the opacity and its wavelength dependence) 
of the outer layers and also back-warming of the innermost ones \citep{gustafsson01}. Changes in the equation of state 
are mainly variations in the mean molecular weight for abundant elements, changes in the number of free electrons for 
elements that are important electron donors, or other more intricate changes related to chemical equilibrium through 
molecule formation. In the next two sections, we give examples of MARCS and ATLAS9 models from our grid, and 
illustrate briefly the changes in the model atmospheres related to large changes in C and $\alpha$ elements. 

\begin{figure}
\includegraphics[width=3.5in,angle=0]{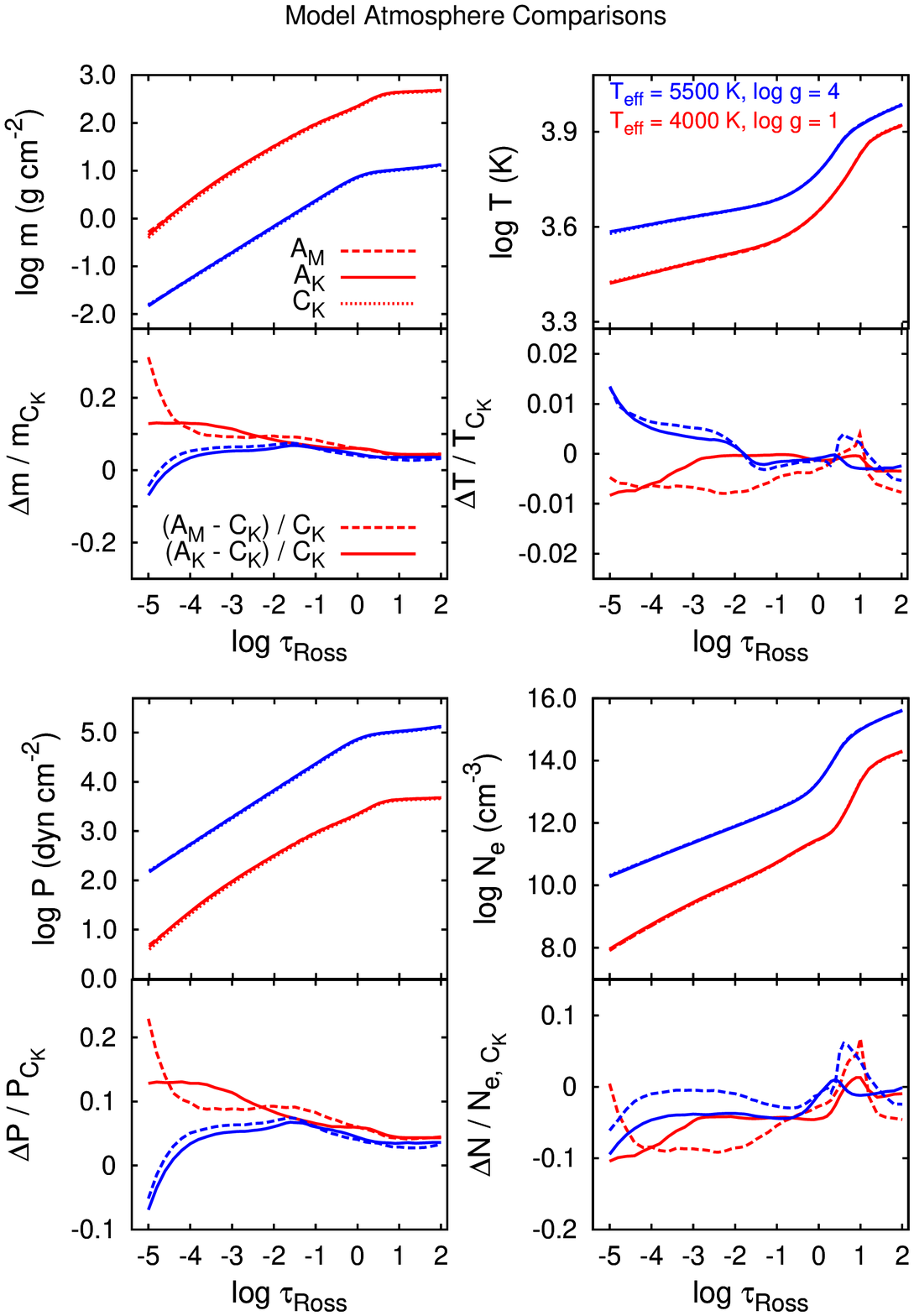}
\caption{Examples of ATLAS9 and MARCS model atmospheres with T$_{\rm eff}$ = 4000\,K and log~$g$ = 1 (red line) 
and T$_{\rm eff}$ = 5500\,K and log~$g$ = 4 (blue line). The model from the APOGEE grid is denoted by a solid line, the
corresponding ATLAS9 model from \citet{castelli01} is denoted by a dotted line, the MARCS model
is denoted by dashed line. The panels show the mass column ($m$), 
temperature ($T$), gas pressure ($P$), electron number density ($N_{\rm e}$) as a function
of the logarithm of optical depth (log $\tau$), and the relative difference between the ATLAS9 
(A$_{\rm K}$), MARCS (A$_{\rm M}$) calculations 
presented in this paper and \citet{castelli01} (C$_{\rm K}$). 
}
\end{figure}

\subsection{Comparing our MARCS and ATLAS9 models to the Castelli-Kurucz grid}

Figure 3 illustrates examples of the atmospheric structures for two models with solar composition, one for 
$T_{\rm eff} = 4000$\,K and log~$g = 1$ (red line) and the other for $T_{\rm eff} = 5500$\,K and log~$g = 4$ (blue line). 
There are four double panels showing the parameter dependence with Rosseland optical depth of the mass column 
(top-left panel), temperature (top-right), gas pressure (bottom-left) and electron number density (bottom-right).
The dashed, solid, and dotted lines correspond to the new APOGEE MARCS, APOGEE Kurucz, and earlier (NEWODF) 
Castelli-Kurucz models \citep{castelli01}, respectively. While the APOGEE Kurucz and MARCS models share
essentially the same chemical composition \citep{asplund01, grevesse02}, the Castelli-Kurucz models use the 
solar mixture given by \citet{grevesse01}.

For both the cooler (red line) and the warmer model (blue line), we see good agreement between the MARCS and the ATLAS9 
models presented in this paper. 
The new MARCS and ATLAS9 models show good agreement, the differences are modest, less than 1 percent, for the 
thermal structure, and less than 2$-$3 percent for the gas pressure and the electron density in the layers where 
weak spectral lines and continuum form ($\tau >$ 0.01). 
These small differences are most likely related to the different equation of state implemented 
in the two codes, and can also be due to the fact that ATLAS9 uses the ODF method, while MARCS uses the 
OS method. The differences between the two increase at temperatures lower than 4000\,K 
due to the different H$_{2}$0 and TiO linelists used in the calculations. 
More importantly, larger differences are present in the gas pressure and electron density in the ATLAS9 models 
compared to the Castelli-Kurucz ones for the cooler models. These differences must be related to the
updated solar chemical composition, as the 
Castelli-Kurucz models are also computed with a corrected H$_{2}$O linelist. The most significant change in the solar 
composition corresponds to the reduction in oxygen, nitrogen and carbon abundances, all of which decrease in the update. 
This causes a reduction in the Rosseland opacity at a given temperature and gas pressure. 
The decreased Rosseland opacity leads to a subsequent increase in the total pressure, which is only 
partially compensated by a reduction in the electron density. 

\citet{gustafsson02} present some further comparisons between MARCS models and ATLAS9 models of \citet{castelli01}. 
Other comparisons between MARCS, ATLAS9, and PHOENIX model atmospheres and spectra were discussed 
by \citet{plez02}. 

\begin{figure}
\includegraphics[width=3.5in,angle=0]{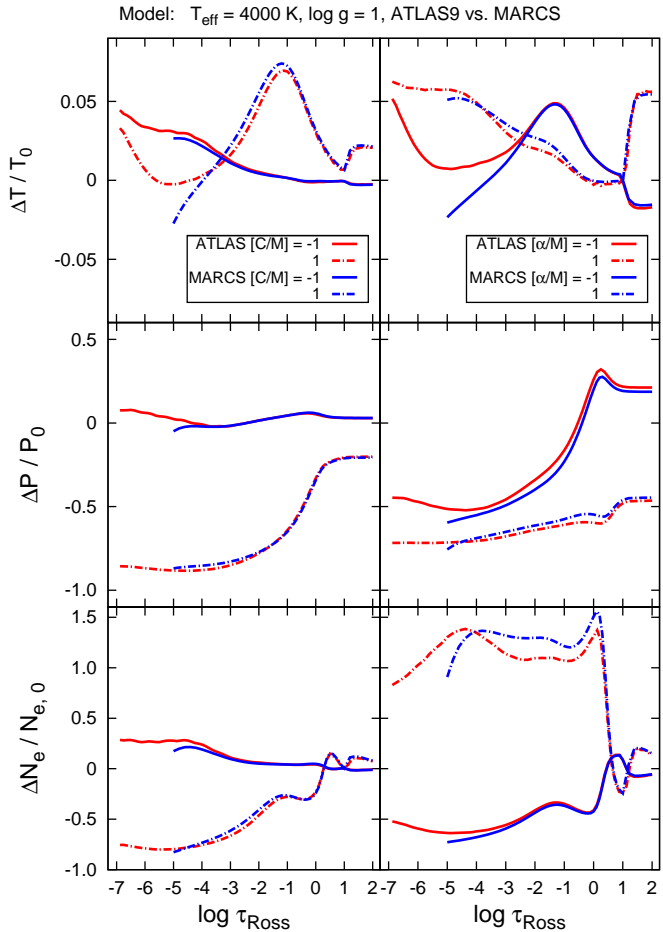}
\caption{Examples of how different [C/M] and [$\alpha$/M] content changes the temperature profile in the ATLAS9 
and MARCS atmospheres for T$_{\rm eff}$ = 4000\,K, log~$g$ = 1.
This figure shows the temperature ($T$), gas pressure ($P$), electron number density ($N_{\rm e}$) and the relative 
difference (where 0.5 corresponds to 50\%, 1 corresponds to 100\%) of these parameters between [C/M]=[$\alpha$/M] = 0
and [C/M]=[$\alpha$/M] = $-$1, and 1 as a function of optical depth (log $\tau$). The symbols 
$T_{0}$, $P_{0}$, $N_{e, 0}$ correspond to [C/M]=[$\alpha$/M] = 0. 
}
\end{figure}

\begin{figure}
\includegraphics[width=3.5in,angle=0]{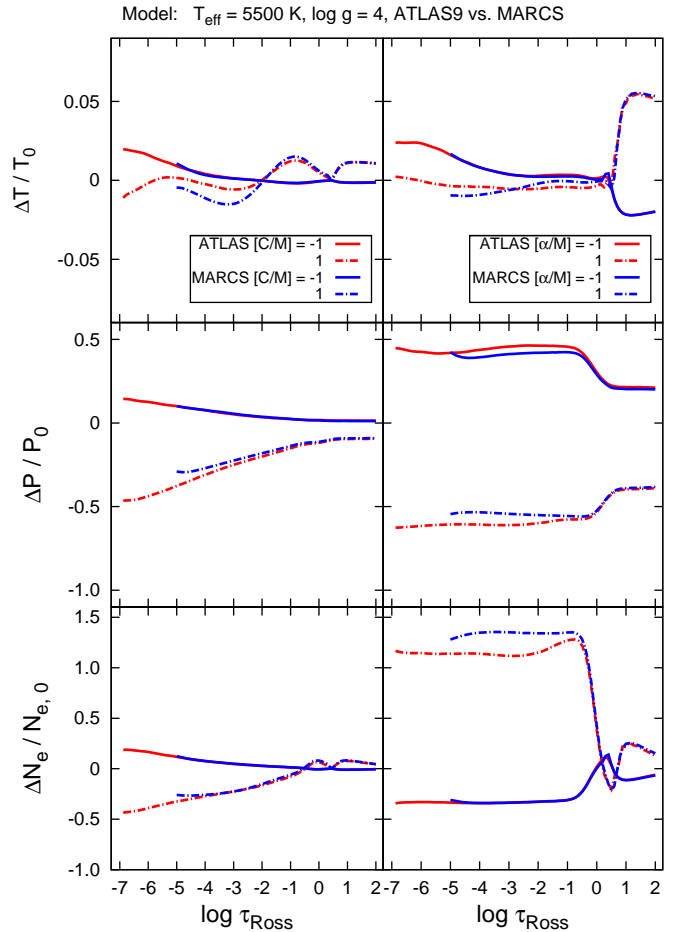}
\caption{Examples of how different [C/M] and [$\alpha$/M] content changes the temperature profile in the ATLAS9 
and MARCS atmospheres for T$_{\rm eff}$ = 5500\,K, log~$g$ = 4.
For more details see the caption of Figure 6.
}
\end{figure}

\subsection{Changes in model structures related to chemical composition}

Figures 4 and 5 illustrate the changes in the model structures in response to changes in 
chemical composition. The left-hand panels of Figure 4 (T$_{\rm eff} = 4000$\,K model) and 5 (T$_{\rm eff} = 5500$\,K) 
show the relative changes associated with variations of the carbon abundance from 
solar proportions. The ATLAS9 model for the warmer temperature has v$_{\rm t}$ = 2 km~s$^{-1}$, while the 
MARCS model has v$_{\rm t}$ = 1 km~s$^{-1}$, but this difference does not affect the model structures significantly. 
The relative variations found are very similar for MARCS and 
ATLAS9, except in the outermost layers of the atmosphere. 
The changes in the thermal structure are modest and show a behavior symmetric to that found for oxygen, which
strongly suggests that CO formation is the driver of the variation. CO is the most tightly bound molecule and 
consumes almost all of the free atoms of either carbon or oxygen, whichever is less abundant, leaving the 
majority species (C I or O I) to form other molecules. If oxygen dominates, it produces a "normal star" 
(if cool, an M star), otherwise carbon is free to form many molecules with high opacities, making carbon 
stars with very different spectra. At T$_{\rm eff} = 5500$\,K, CO formation is low, thus this molecule does not affect the 
chemical equilibrium contrary to what is seen in the cooler atmospheres. 

The right-hand panels show changes for the same model parameters and different $\alpha$-element abundances. 
Large changes are visible for both models in the pressure and electron numbers. The significant differences in the 
pressure compared to the cooler model are due to the increased gravity. Increasing the abundance of $\alpha$ 
elements reduces pressure (despite the fact that electron pressure increases), because the electrons of the other 
main electron contributors change the continuum opacity.
The $\alpha$ elements that contribute most to the total number of electrons in the ATLAS9 models are shown in Figure 6. 
For the T$_{\rm eff} = 4000$\,K, log~$g = 1$ model, the main electron contributors are Ca, Mg, Na, and Al 
in the outer layers and Mg, Si and Fe in the deeper layers. For the warmer models, this changes significantly, as more 
Fe and H is ionized, the overall number of electrons increases and the main 
electron contributors become Mg, Si, and Fe through most of the atmospheres. 

Internal tests showed that polyatomic carbon molecules (C$_{2}$H$_{2}$, C3) substantially change the structure of the 
atmosphere with C/O ratios higher 1.7, if they are included in the linelist. These molecules significantly inflate the
atmosphere changing the temperature in the line-forming photospheric layers. Since these molecules were used neither in 
the ATLAS9, nor in the MARCS calculations, atmospheric structures with C/O~$>$~1.7 and T$_{\rm eff} < 4000$\,K are not 
reliable and thus not published.

\begin{figure}
\includegraphics[width=3.5in,angle=0]{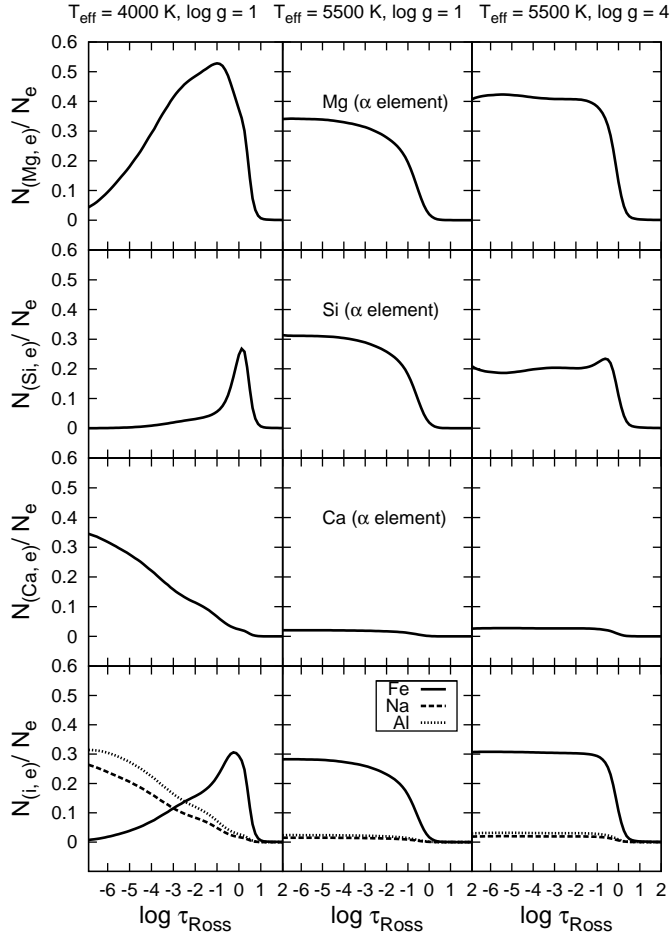}
\caption{Examples of the largest electron contributors to the total electron numbers in the ATLAS9 atmospheres 
for [M/H]=[C/M]=[$\alpha$/M] = 0. The figure shows the element electron numbers relative to the total number of
electrons for the two models used in previous figures. The T$_{\rm eff} = 5500$\,K, log~$g = 1$ model is plotted 
in the middle section to show that the electron contributors do not change significantly compared to a model 
with the same temperature but higher gravity. 
}
\end{figure}

\section{Conclusions}

Most of the ATLAS9 models are fully converged 
above T$_{\rm eff}$ = 5000\,K. Convergence problems are visible only in the 
outermost layers for stars with T$_{\rm eff} <$ 5000\,K. The regions affected by convergence issues are limited to 
log $\tau_{\rm Ross} < -$4. However, these un-converged layers on the top of the atmosphere at low temperatures 
do not affect most line profiles very significantly, because most lines form deeper in the atmosphere. A little over a
million ATLAS9 atmospheres were calculated. 
 
Over 20,000 MARCS fully converged models with ranges in [C/M] and [$\alpha$/M] have been produced. 
Models are available in spherical and plane-parallel cases. Convergence issues are also present for red dwarf stars, 
similarly to the ATLAS9 models. 

Examples were presented for both ATLAS9 and MARCS models for T$_{\rm eff} = 4000$\,K and 5500\,K, and compared to the 
Castelli-Kurucz grid. The new ATLAS9 and MARCS models agree well for all temperatures, while the differences between these
calculations and the Castelli-Kurucz grid arise 
from the updated abundance tables. We briefly illustrated the effects of decreased/increased carbon and $\alpha$ content 
on the structure of the atmospheres, which are very similar in the MARCS and ATLAS9 models. 
The response to the carbon content is only different in the outermost layers due to increased CO in the atmosphere. 
The response to $\alpha$ elements is also almost the same in MARCS and 
ATLAS9 models. The increased $\alpha-$element content has profound effects in electron numbers and pressure 
for both the giant and dwarf stars, which is related to the higher number of Mg, Si, and Ca ions. 
Carbon rich models with C/O~$>$~1.7 and T$_{\rm eff} < 4000$\,K are not published here, because polyatomic carbon
molecules not included in the linelists here significantly change the temperature structure in the photosphere. 

These model grids will be used as the primary database for the pipeline analysing the spectra from the APOGEE survey. 
The high-resolution model spectra used in the 
survey's atmospheric parameters and abundances determination code will be built on the models 
presented in this paper. Both the ATLAS9 and MARCS models atmospheres will be continuously updated with new 
compositions as the APOGEE survey progresses. The calculated ODFs, Rosseland opacities available for 
v$_{\rm t}$ = 0, 1, 2, 4, 8 km~s$^{-1}$, and the ATLAS9 model atmosphere files are available for 
v$_{\rm t}$ = 2 km~s$^{-1}$ from the ATLAS-APOGEE website\footnote{http://www.iac.es/proyecto/ATLAS-APOGEE/}. The MARCS
models are available with v$_{\rm t}$ = 2 km~s$^{-1}$ for the giant and v$_{\rm t}$ = 1 km~s$^{-1}$ for the dwarf stars 
from the standard MARCS website\footnote{http://marcs.astro.uu.se/}.

\acknowledgements{We are grateful to Robert L. Kurucz, who kindly provided us with the new and corrected H$_{2}$O linelist. 
We thank Katia Cunha and Verne Smith for helping us with the discussion of Kurucz and 
MARCS models. We would also like to thank Martin Asplund and Kjell Eriksson for their contributions to the 
MARCS models presented in this paper. Don P. Schneider, Kaike Pan, and David H. Weinberg also greatly improved our 
work by providing many useful comments and suggestions. 

Funding for SDSS-III has been provided by the Alfred P. Sloan Foundation, the Participating Institutions, the 
National Science Foundation, and the U.S. Department of Energy Office of Science. The SDSS-III web site is 
http://www.sdss3.org/.

SDSS-III is managed by the Astrophysical Research Consortium for the Participating Institutions of the SDSS-III 
Collaboration including the University of Arizona, the Brazilian Participation Group, Brookhaven National Laboratory, 
University of Cambridge, Carnegie Mellon University, University of Florida, the French Participation Group, the 
German Participation Group, Harvard University, the Instituto de Astrofisica de Canarias, the 
Michigan State/Notre Dame/JINA Participation Group, Johns Hopkins University, Lawrence Berkeley National Laboratory, 
Max Planck Institute for Astrophysics, New Mexico State University, New York University, Ohio State University, 
Pennsylvania State University, University of Portsmouth, Princeton University, the Spanish Participation Group, 
University of Tokyo, University of Utah, Vanderbilt University, University of Virginia, University of Washington, 
and Yale University. 
}

\end{document}